\documentclass[reprint, amsmath,amssymb,aps,superscriptaddress,twocolumn]{revtex4-1}
\usepackage{graphicx}
\usepackage{dcolumn}
\usepackage{bm}
\usepackage{color}      
\usepackage{setspace}
\usepackage{subfigure}
\usepackage[plainpages=false, pdfpagelabels,
    pdfpagemode=none,
    pdfstartview=FitH,
    colorlinks=true,
    linkcolor=blue,
    citecolor=blue,
    a4paper=true,
    pagebackref=false,
    pdfstartpage=1,
    pdfauthor={F.Mirjani},
    pdfsubject={Notes},
    pdftitle={},
]{hyperref}
\begin{document}
\preprint{APS/123-QED}
\title{Opportunities and limitations of transition voltage spectroscopy: a theoretical analysis}
\author{F. Mirjani}
\author{J. M. Thijssen}
\affiliation{Kavli Institute of Nanoscience, Delft University of Technology, Lorentzweg 1, 2628 CJ Delft, The Netherlands}
\author{S. J. van der Molen}
\affiliation{Kamerlingh Onnes Laboratorium, Leiden University, Niels Bohrweg 2, 2300 RA Leiden, The Netherlands}
\date{\today}

\begin{abstract}
In molecular charge transport, transition voltage spectroscopy (TVS) holds the promise that molecular energy levels can be explored at bias voltages lower
than required for resonant tunneling. We investigate the theoretical basis of this novel tool, using a generic model. In particular, we study the length dependence of the conducting frontier orbital and of the `transition voltage' as a function of length. We show that this dependence is influenced by the amount of screening of the electrons in the molecule, which determines the voltage drop to be located at the contacts or across the entire molecule. We observe that the transition voltage depends significantly on the length, but that the ratio between the transition voltage and the conducting frontier orbital is approximately constant only in strongly screening (conjugated) molecules. Uncertainty about the screening within a molecule thus limits the predictive power of TVS. We furthermore argue that the relative length independence of the transition voltage for non-conjugated chains is due to strong localization of the frontier orbitals on the end groups ensuring
binding of the rods to the metallic contacts. Finally, we investigate the characteristics of TVS in asymmetric molecular junctions. If a single level dominates the transport properties, TVS can provide a good estimate for both the level position and the degree of junction asymmetry. If more levels are involved the applicability of TVS becomes limited.

\begin{description}
\item[PACS numbers]
\end{description}
\end{abstract}
\maketitle

\section{Introduction}
\begin{figure}
\includegraphics[scale=0.35]{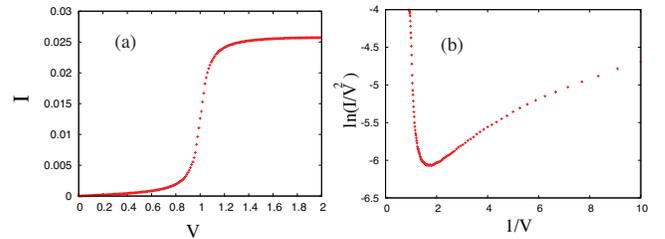}
\caption{\label{fig:tvs} (a) Typical current vs. the bias voltage curve for a model calculation (see main text). Once a molecular level is aligned to the
Fermi energy of one of the electrodes, a step appears (at $V=1$ V in this example where the molecular level energy is $E=0.5$ , $E_F=0$). (b) Fowler-Nordheim (FN) plot created from (a), showing $\ln(I/V^2)$ vs. $1/V$. A minimum appears at a voltage $(V_{\text{min}})$, which is smaller than the voltage required to reach the molecular level. \\ }
\end{figure}

Molecular electronics aims at investigating and exploring the quantum properties of molecules in electronic devices \cite{Nitzan, Datta, Cuevas, Molen}. An essential parameter when considering charge transport through molecules is the location of the molecular energy levels. In general, these levels are a few electron Volts $(eV)$ away from the Fermi energy ($E_F$) of the electrodes. Hence, it should in principle be possible to align the Fermi energy of one of the electrodes with a molecular energy level by applying a bias voltage. However, in practice a junction often breaks down before the molecular level is reached, as a result of the gigantic electric field coming about ($\sim 10^9 V/m$).\\
Recently, Beebe \textit{et al.} introduced transition voltage spectroscopy (TVS) as an alternative method to characterize molecular energy levels in a device geometry \cite{Beebe,Beebe2}. They determined current $(I)$ vs. voltage $(V)$ characteristics for a series of molecular devices, and replotted their data in a Fowler-Nordheim (FN) manner \cite{Fowler}, i.e.\ they plotted $\ln(I/V^2)$ versus $1/V$. In such graphs a clear minimum appears, at a voltage, $V_{\text{min}}$, which is generally \textit{smaller} than the voltage needed to reach the molecular level (see Fig.~\ref{fig:tvs} for a calculated example). Beebe \textit{et al.} proposed that $V_{\text{min}}$ provides direct information about the energy distance between $E_F$ and the nearest molecular level. This interpretation was based on a picture of molecular junctions as tunnel barriers obeying the Simmons model for charge transport \cite{Simmons}. Experimentally, their proposition was supported by an extended series of experiments \cite{Beebe,Beebe2}. They demonstrated that $V_{\text{min}}$ does not vary with molecular length $d$ for alkanethiols, which indeed have a length-independent HOMO-LUMO gap (The terms HOMO and LUMO refer to the highest occupied molecular orbital and lowest unoccupied molecular orbital, respectively). Furthermore, they experimentally showed that for conjugated molecules $V_{\text{min}}$ scales linearly with $|E_F-E_{\text{HOMO/LUMO}}|$, depending on which level, HOMO or LUMO, is closest to the Fermi energy.\\
Recently, Huisman \textit{et al.} found that the Simmons picture is inconsistent with the experimental data of Beebe \textit{et al.} \cite{Huisman}. They showed that within that model, $V_{\text{min}}$ is expected to decrease like $1/d$ for a series of alkanethiols. (If the image potential is taken into account, this functional dependence on $d$ changes for small $d$, see also Ref. \cite{Trouwborst}). Within a coherent molecular transport model, however $V_{\text{min}}$ was found to be length independent for the same molecular series, provided $d> 8 \mathring{A}$ \cite{Huisman}. Additionally, both Huisman and Araidai \textit{et al.} pointed out that $V_{\text{min}}$ does not occur at the transition between direct tunneling and FN tunneling \cite{Huisman, Tsukada}. Within a molecular level model, $V_{\text{min}}$ rather appears when a certain amount of the tail of the broadened resonant level has come within the bias window. In two recent papers, the Thygesen group took the discussion a step further, by performing \textit{ab initio} calculations for a set of molecular junctions \cite{Chen, Markussen}. In particular, they pointed out that junction asymmetry is an essential parameter for TVS to be interpreted correctly \cite{Chen}.\\
TVS has clear potential in analyzing molecular charge transport experiments. However, the very issue if it can indeed be applied generally is still not settled. For this reason, we present an investigation of TVS using a generic theoretical model. Our analysis relies on a DFT-based many-body method described recently \cite{Mirjani}. We focus on two essential questions that were brought up recently. The first one follows directly from the work of Huisman \textit{et al.} \cite{Huisman}. An essential difference between the two models they compared is in the voltage profile assumed. In their coherent molecular level picture, the full voltage drops at the contacts, whereas in the Simmons-based model, the potential decreases linearly over the molecule itself. Hence, it is not clear whether the distinction in length dependence of $V_{\text{min}}$ is due to the voltage profile or due to the other clear differences between these models. Here, we study the influence of the exact voltage profile for a generic molecular model. The second issue we investigate involves the consequence of asymmetry for TVS (see Ref. \cite{Chen}). Specifically, we introduce two separate $V_{\text{min}}$ values, for both positive and negative bias. Subsequently, we study how (and within which conditions) these quantities are related to both the position of the nearest molecular level(s) and the junction asymmetry itself. \\
The organization of this paper is as follows. In Sec.~\ref{Model} we introduce our model. As a generic system we use a Hubbard chain connected to two non-interacting conducting leads. To investigate the conductance through the molecule, we combine local spin density approximation (LSDA) with many-body Green's functions. In Sec.~\ref{drop}, the effect of a voltage drop over a molecule is studied, whereas the role of (a)symmetry is discussed in Sec.~\ref{symmetry}.

\section{Model and method}
\label{Model} 
\begin{figure}
\includegraphics[scale=0.7]{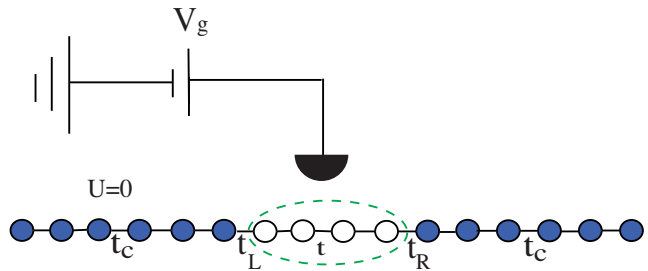}
\caption{\label{fig:model}A short interacting Hubbard chain connected to two non-interacting leads. The gate voltage $V_g$ can be applied to the interacting region. The hopping terms in the interacting part and in the contacts are $t$ and $t_c$ respectively. The interacting region is coupled to the left and right contact by $t_L$ and $t_R$.   \\   }
\end{figure}

The system we consider consists of a small region where possibly Coulomb interactions are present, weakly coupled to two non-interacting, semi-infinite leads (see Fig.~\ref{fig:model}). The interacting region contains one or several quantum dots in series. The Hamiltonian of the leads and the central region read respectively (H.c. denotes the Hermitian conjugate)
\begin{equation}
H_{\text{leads}}=-t_c\sum_{\eta=L,R}  \sum_{\sigma} \sum_ {i} [c^\dagger_{i,\eta,\sigma} c_{i+1,\eta,\sigma}+\text {H.c.}]
\end{equation}
and
\begin{eqnarray}
&\displaystyle H_{\text{molecule}} =-t \sum_{\sigma} \sum_ {i=1}^{N_L-1} [d^\dagger_{i,\sigma} d_{i+1, \sigma}+\text {H.c.}] \nonumber \\
&\displaystyle + U \sum_{i=1}^{N_L} d^{\dagger}_{i \uparrow} d_{i \uparrow} d^{\dagger}_{i \downarrow}  d_{i \downarrow}+\epsilon \sum_{\sigma} \sum_{ i=1}^{N_L}  d^{\dagger}_{i \sigma} d_{i \sigma}
\label{HM2}
\end{eqnarray}
where $N_L$ is the length of the interacting chain. The parameters $t$ and $t_c$ represent the hopping rate in the molecule and contacts respectively, and $U$ describes the on-site Coulomb interaction. However, since we discuss the off-resonant regime and since in TVS the first step in the I-V characteristic is dominant, we can put $U=0$ without loss of generality. The creation and annihilation operators, $c^\dagger_{i}$, $d^\dagger_{i}$, $c_{i}$ and $d_{i}$ acting on site $i$ satisfy the usual anti-commutation relations. In addition, the external gate potential, $V_g$, can be applied to the central region which is included in the energy, i.e. $\epsilon=\epsilon(V_g)$. The index $\sigma=\uparrow,\downarrow$ describes the spin.\\
The coupling Hamiltonian between the molecule and the contacts reads
\begin{equation}
H_{\text{coupling}}= \sum_{\substack{\eta=L,R \\ \sigma }} [t_{\eta}c^\dagger_{i,\eta,\sigma} d_{j,\sigma}+\text {H.c.}]
\end{equation}
where $i$ denotes the leftmost (rightmost) site of the right (left) contact and $j$ corresponds to the leftmost (rightmost) site of the central region.\\
Our method is based on a mapping of the Hamiltonian of the central region to a limited set of many-body eigenstates. All of the parameters in the many-body energy spectrum are obtained from ground state L(S)DA calculations \cite{Capelle}. We then calculate the transport using many-body Green's function theory (see Ref. \cite{Mirjani} for details).
The retarded Green's function of a single level (without any Coulomb interaction) connected to the electrodes is described by
\begin{eqnarray}
 G^r (\omega)= \frac{1}{\omega-E_s-\Sigma^{r}}
\end{eqnarray}
where $E_s$ is the single electron energy \cite{Haug}. The presence of the contacts is taken into account by the retarded self-energy, $\Sigma^r$. It is necessary to calculate the effective coupling $(t^{\text{eff}})$ of a level $\alpha$ to the contacts. We do this by projecting the central chain Hamiltonian onto two many-body states of $N$ and $N+1$ or $N+1$ and $N+2$ \cite{Mirjani}. By extracting $t^{\text{eff}}$, the calculation of the self-energies is straightforward.  Once the Green's functions are known, the current can be calculated from a Landauer type of equation
\begin{eqnarray}
&\displaystyle I=\frac{ie}{2h}\int   \,d\omega   \text{Tr}\{ (\Gamma_L(\omega)-\Gamma_R(\omega))G^<(\omega)+ \nonumber \\
&\displaystyle (f(\omega,\mu_L)\Gamma_L(\omega)-f(\omega,\mu_R)\Gamma_R(\omega))(G^r(\omega)-G^a(\omega) )   \}  
\end{eqnarray}
where $G^<$ and $G^a$ are the lesser and advanced Green's functions respectively, $\Gamma_j=i(\Sigma_j^r- \Sigma_j^{r^{\dagger}})$ and $f(\omega,\mu_j)$ is the Fermi distribution of lead $j$.\\
In particular, for the transport through a single level or if the left and right line-width functions are proportional to each other, i.e. $\Gamma_L=\lambda \Gamma_R$, the current can be written as
\begin{equation}
I=\frac{ie}{h}\int \,d\omega  \text{Tr}\{ \frac{\Gamma_L\Gamma_R}{\Gamma_L+\Gamma_R}(G^r-G^a)\} 
(f(\omega,\mu_L)-f(\omega,\mu_R)) 
\end{equation}

\section{Results and discussion}
\subsection{Length dependence of $V_{\text{min}}$ }
\label{drop}
\begin{figure}
\includegraphics[scale=0.6]{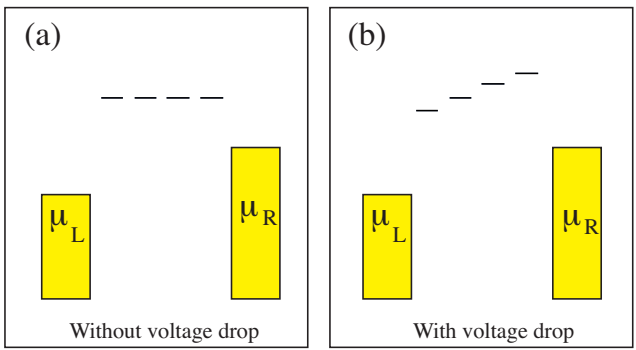}
\caption{\label{fig:drop-modified}Schematic representation of the energy levels of a short Hubbard chain connected to two non-interacting leads (with symmetrically applied bias voltage) considered in two different configurations. (a) without any voltage drop on the central region: the voltage drops only at contacts. (b) The potential decreases linearly with distance. \\ }
\end{figure}
In the Simmons model, $V_{\text{min}}$ is found to be inversely proportional to the molecular length while in the molecular model of Ref. \cite{Huisman}, $V_{\text{min}}$ is found to be independent of the molecular length for $d> 8 \mathring{A}$. Huisman \textit{et al.} mentioned that the differences in the functional dependence of $V_{\text{min}}$ on the length of the molecule could originate from the different voltage profiles in these two models.\\
In this section, we investigate the length dependence of $V_{\text{min}}$ in more detail. Different factors influence the length dependence of transition voltage: screening, the hopping integral and the spatial structure of the conducting orbitals. Here we systematically discuss the influence of these factors.

\subsubsection{Screening effect and hopping integral}
\begin{figure}
\includegraphics[scale=0.25]{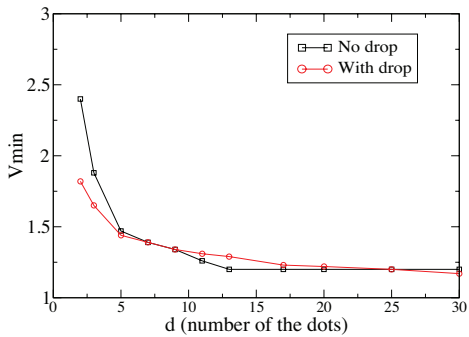}
\caption{\label{fig:d-Vm} $V_{\text{min}}$ versus the length of the molecule (the number of the dots in the central region). $t=1$, $t_c=3$, $U=0$ and $V_g=-3$. Squares for the model (a) of Fig.~\ref{fig:drop-modified} and circles for the model (b) of Fig.~\ref{fig:drop-modified}. \\  }
\end{figure}

\begin{figure}
\includegraphics[scale=0.4]{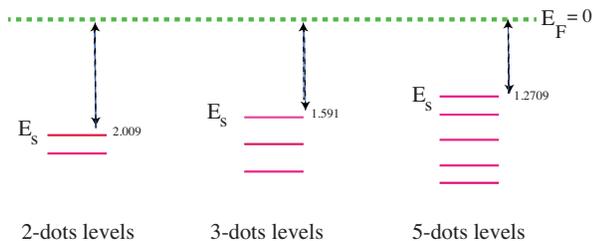}
\caption{\label{fig:molecular-length} Schematic representation for variation of $|E_s-E_F|$ versus the length of the molecule. By increasing the length of
the molecule (the number of the dots in the central region) the closest level to $E_F$ becomes closer to $E_F$ and thus $V_{\text{min}}$ is decreased.
$t=1$, $U=0$ and $V_g=-3$. \\   }
\end{figure}

\begin{figure}
\includegraphics[scale=0.25]{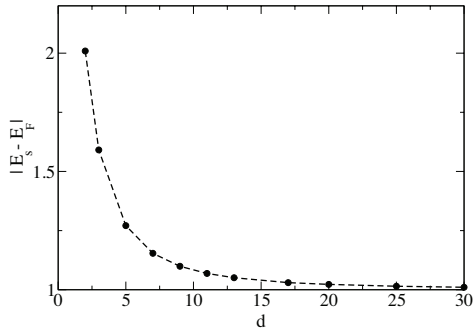}
\caption{\label{fig:Es-Ef-d} $|E_F-E_s|$ versus the length of the molecule (the number of the dots in the central region) for the case that the voltage
drops only at the contacts. $t=1$, $U=0$ and $V_g=-3$. \\}
\end{figure}

The capability of the electrons to screen out the field determines the way in which the voltage drops over the molecule. To investigate the difference between a voltage drop over the contacts and a linear drop over the entire molecule, we use our generic model and consider the two configurations shown in Fig.~\ref{fig:drop-modified}. Fig.~\ref{fig:drop-modified}(a) corresponds to the case of a molecular junction at nonzero bias with the drop entirely over the contacts. Fig.~\ref{fig:drop-modified}(b) corresponds to the case where the voltage decreases linearly with distance between the leads. For model (b), a voltage dependent term is added to the diagonal elements of the Hamiltonian. This term equals $E_F-V/2+(i-1)V/(N_L-1)$ for site $i$, where $N_L$ is the number of central dots. In both cases (a) and (b), the hopping term in the central region is $t$. We note that in Ref.~\cite{Tsukada} a comparable model is presented. However, in that paper, two cases were studied in which not only the voltage drop was different, but also the hopping rate $t$ and the energy values of the molecular levels were chosen differently, making it difficult to distinguish between the individual effects. Here we want to analyze the behavior in a more general context by additionally calculating $V_{\text{min}}$ versus molecular length. \\
First of all, for both models (a) and (b), we indeed see a minimum in the FN plot which is consistent with the results of Ref.~\cite{Tsukada}, confirming that TVS should not be necessarily interpreted as a transition between rectangular and trapezoidal barriers. The results of $V_{\text{min}}$ versus $d$ for models (a) and (b) of Fig.~\ref{fig:drop-modified} are shown in Fig.~\ref{fig:d-Vm}. This plot shows a very similar behavior for the two, rather different, models (a) and (b). Both curves decay to become approximately constant only beyond a length of about 15 dots \cite{notemujica}.\\
It is useful to study, in addition to the $d$-dependence of $V_{\text{min}}$, the dependence on $d$ of the parameter 
\begin{equation}
\chi = \frac {|E_F-E_s|} {V_{\text{min}} },
\end{equation}
suggested in Ref. \cite{Chen}. This parameter gives an indication of the relation between the TVS minimum and the nearest resonance, removing a possible length dependence of the latter from the problem (see below).\\ 
We have plotted $\chi$ versus $d$ in Fig.~\ref{fig:d-Es-Vm}. We see that there is a significant difference between length dependence of $\chi$ for a voltage drop only at the contacts versus a drop over the entire molecule. It should be noted that the main differences between the curves in Fig.~\ref{fig:d-Vm} (rather similar for model (a) and (b)) and Fig.~\ref{fig:d-Es-Vm} (which shows a striking difference between these two models) persist for different values of $t$ and $V_g$.\\
In fact the parameter $\chi$ is a key quantity in our analysis as it gives more insight into the performance of TVS than $V_{\text{min}}$. Moreover, if $\chi$ is a well-defined number, the gap between the Fermi level and the nearest molecular level is easily calculated from $V_{\text{min}}$. Returning to Fig.~\ref{fig:d-Es-Vm}, we see that the parameter $\chi$ shows the expected difference between voltage drop over the entire molecule versus a drop over the contacts only, whereas inspection of $V_{\text{min}}$ versus length, may not enable us to distinguish clearly between the two cases, as the HOMO or LUMO level itself may vary significantly with length. \\
Two factors influence the behavior of the quantities $V_{\text{min}}$ and $\chi$ versus length: the extent to which the electrons screen out the applied potential, and the hopping integral $t$. In Ref.~\cite{Tsukada} it was pointed out that the two are related: strong screening usually implies a large value of the hopping integral, as both result from a high mobility of the electrons. We discuss both effects in detail.\\
\begin{figure}
\includegraphics[scale=0.25]{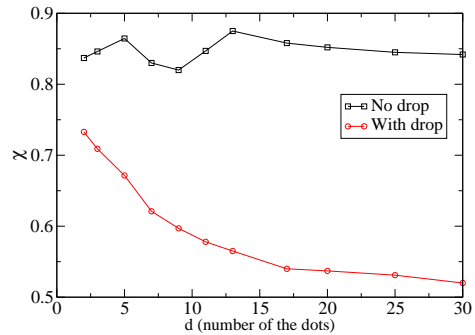}
\caption{\label{fig:d-Es-Vm} $\chi = |E_F-E_s|/V_{\text{min}}$ versus the length of the molecule (the number of the dots in the central region). $t=1$,
$t_c=3$, $U=0$ and $V_g=-3$. Squares for the model (a) of Fig.~\ref{fig:drop-modified} and circles for the model (b) of Fig.~\ref{fig:drop-modified}. \\
}
\end{figure}
The screening length in molecular systems varies strongly across different molecules due to the characteristics of different chemical bonds. For instance, screening is usually strong when there are many $\pi$ electrons in the molecule, in particular when they are arranged along conjugated pathways \cite{RuiLiu, Liang}. The amount of screening is related to the HOMO-LUMO gap: A small HOMO-LUMO gap is indicative of strong screening \cite{Nitzan1}. Screening strongly influences the energy landscape through which the electrons move: in general if the electrons have the ability to screen out the voltage, the voltage drop will be localized near the contacts, while in the opposite limit the voltage drops over the entire molecule. As we have seen, the difference between strong and weak screening is most clearly observed in the dependence of $\chi$ on the length, and very weakly in $V_{\text{min}}$. \\
On the other hand, the hopping integral $t$ causes the HOMO and LUMO levels to vary with the chain length. In particular, for a chain of length $d$, the relation between the frontier orbital and $t$ can be seen in the energy spectrum given by
\begin{equation}
\label{jalab}
E_n-E_0=-2t\cos (ka)
\end{equation}
where $a$ is the inter-site distance (which is $1$ in our case) and $k=n\pi/L$ where $L=(d+1) a$ and $E_0$ is the energy offset. Thus the maximum width of the energy spectrum is $4t$, and the minimum energy for a chain of length $d$ (if it were isolated) is given by 
\begin{equation}
E_1 = E_0 - 2t\cos\left[\pi/(d+1)\right].
\end{equation}
This level is relevant for LUMO transport; for HOMO transport the relevant level would be $E_0 + 2t\cos\left[\pi/(d+1)\right]$. We denote the energy level closest 
to the Fermi energy of the leads (the \emph{frontier orbital}) by $E_s$. As shown schematically in Fig.~\ref{fig:molecular-length}, when increasing the molecular length, the dominant transport level $(E_s)$, moves closer to the Fermi energy (where in this example all of the levels are below the Fermi energy). Fig.~\ref{fig:Es-Ef-d} shows the value $E_s$ of the frontier orbital with respect to the Fermi energy as a function of chain length. In fact, this variation of $E_s$ with length, is the dominant factor in Fig.~\ref{fig:d-Vm}. \\
To judge the effect of the hopping integral in a real molecule we consider known values for this parameter for the case of conjugated and non-conjugated molecules. The hopping integral $t$ is reported to be 3.18 (eV) for a phenyl ring and 1.68 (eV) for alkyl \cite{Kohler}. Although these values show the expected trend, their difference does not seem dramatic enough to be responsible for the substantially different conductance of the two species. It seems therefore that the major difference between the conductance properties of conjugated and non-conjugated systems is not so much due to the difference in hopping integral. These two classes of molecules show however quite a different HOMO-LUMO gap, which indicates that the amount of screening is much larger in the conjugated case. \\
We conclude that realistic values of the hopping integral lead to a $V_{\text{min}}$ which decays with length. This length dependence is expected to be noticeable in both conjugated and non-conjugated molecules. When looking at the parameter $\chi$, this length dependence disappears when the screening effects in the molecule are strong. In the next subsection we investigate why the length dependence of $V_{\text{min}}$ observed in the experiment is weak \cite{Beebe2}. \\

\begin{figure}
\includegraphics[scale=0.25]{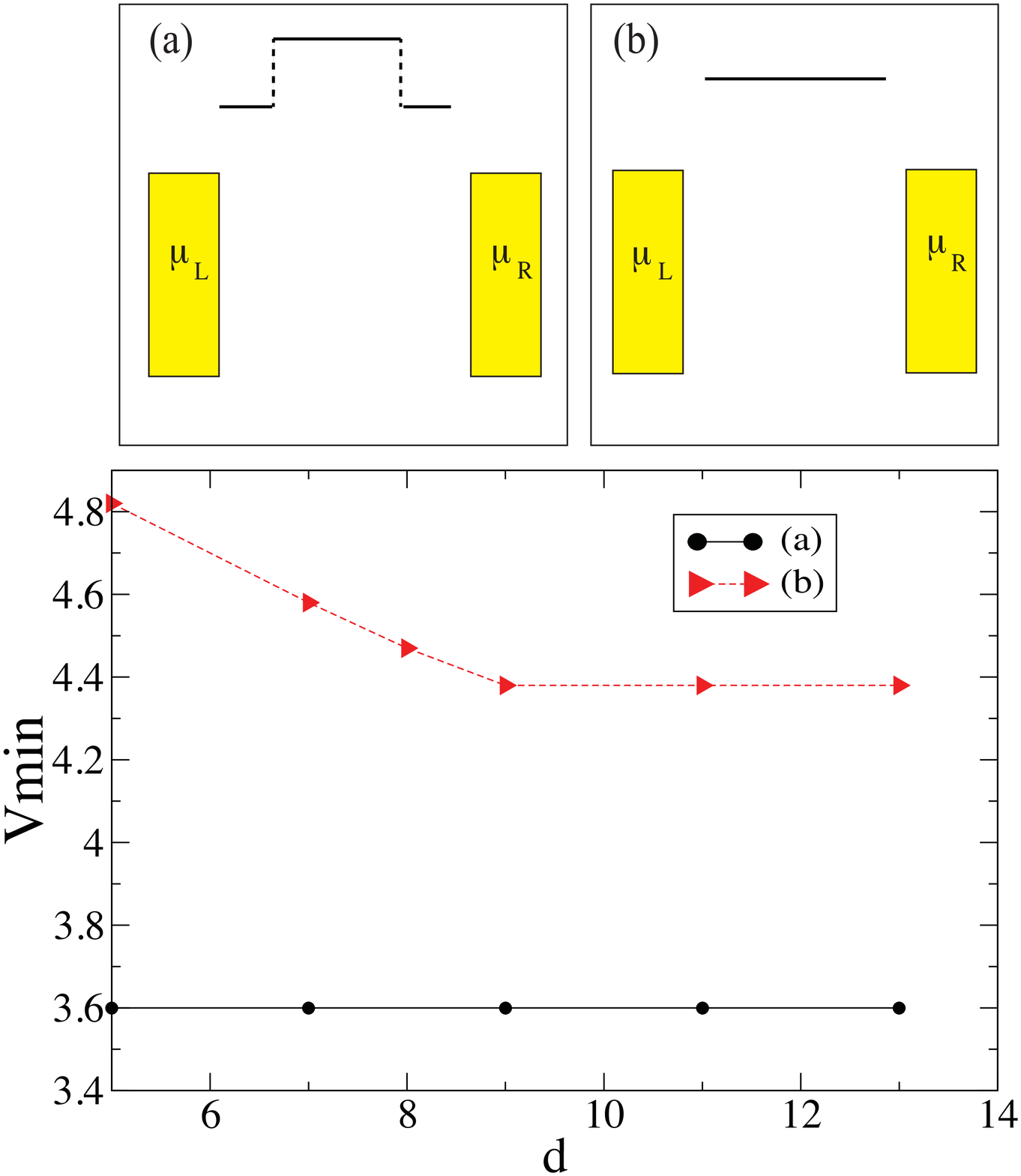}
\caption{\label{fig:uuu} The hopping integral of an alkanethiol system is almost 1.5 eV \cite{Kohler}. Moreover DFT calculations show that the energy difference between frontier level of alkane located on sulphur and the rest of the alkane chain is about $E_a=1.5$ eV. Considering $E_{\text{frontier}}=3.5$ and $E_a=1.5$, it is possible to calculate $E_0$ from Eq.~\ref{jalab} for the alkane chain which is about 8 eV. Then the parameters $E_0$ and $t$ can be utilized to investigate the transport through such a chain using our generic model. Here have shown $V_{\text{min}}$ versus the length of the molecule for alkane chains with those parameters for (a) a Hubbard chain in which the first and last site are at a chemical potential of $3.5$ eV and the rest of the chain is at the chemical potential of $8$ eV and (b) a chain where all of the sites are at a chemical potential of $6.5$ eV. $t=1.5$, $t_c=5$ . \\}
\end{figure}
\subsubsection{Spatial structure of frontier orbitals}
We have seen that $V_{\text{min}}$ varies with $d$ for $d < 15$, irrespective of the voltage drop. However, Beebe \textit{et al.} concluded from an extensive analysis of experiments that $V_{\text{min}}$ varies much less with $d$ in alkanethiols \cite{Beebe2}. This is also shown by H. Song \textit{et al.} in Ref. \cite{Reed}. We argue that the reason for this lies in the fact that the transport orbital is located mainly on the sulphur binding group \cite{Evers, Li1, Li2, Onipko1, Kaun, Liu}. Also, for the series of conjugated molecules used by Beebe \textit{et al.} \cite{Beebe, Beebe2}, DFT calculations \cite{nw} suggest that the frontier orbitals or levels close in energy to the frontier orbitals are distributed over the entire molecule (in contrast to the alkanethiols where the transport orbitals are localized on the binding group). This may explain the length dependence of these molecules and why, for larger $t$, this length dependence should be stronger as HOMO/ LUMO levels vary stronger. \\
To study the influence of the localization in alkanethiols, we include two sites which are located at the ends of the molecular chain of our model. To be specific, we consider a Hubbard chain in which the first and last site are at a chemical potential of $3.5$ eV and the rest of the chain is at the chemical potential of $8$ eV and we compare this case with a chain where all of the sites have the same chemical potential of $6.5$ eV. These numbers come from the DFT calculations for alkanethiols where the energy of the frontier orbital is about 3.5 eV and the energy difference between the frontier level of alkane located on sulphur and the rest of the alkane chain is about $1.5$ eV \cite{nw}. Incorporating the energy level value with $2t=3$, $E_0$ for the first case ( (a) in Fig.~\ref{fig:uuu}) would be about 8 eV and for the second case ( (b) in Fig.~\ref{fig:uuu}) is 6.5. These numbers can also be seen from Eq.~\ref{jalab}. The results are shown in Fig.~\ref{fig:uuu}. $V_{\text{min}}$ for the first case is almost length-independent while for the second case, $V_{\text{min}}$ varies significantly with $d$ for $d < 9$.\\
In a recent paper \cite{zCheng}, it has been shown that a direct coupling (i.e. without a sulphur or other binding atom) between a carbon atom and gold is possible, with a high conductance when the carbon is $\text{sp3}$ hybridized. In that case, $V_{\text{min}}$ may vary more strongly with length. So far, TVS results for these alkanes have not been published.\\
All in all, we conclude that for a homogeneous chain, the voltage drop has only a modest direct influence on the dependence of $V_{\text{min}}$ on $d$. Comparing the variation of $\chi = \frac {|E_F-E_s|} {V_{\text{min}} }$ as a function of $d$, we observe however a strong difference between a voltage drop over the contacts or across the molecule. For molecules with strong screening, the HOMO and LUMO are usually located close to the Fermi energy, and this increases the relative variation of $V_{\text{min}}$ with $d$. Finally, the weak length dependence of $V_{\text{min}}$ for non-conjugated molecules must be accounted for by the strong localization of the orbitals on the end group.\\
Finally, from Fig.~\ref{fig:d-Es-Vm}, we infer that for molecules with strong screening, $\chi$ is a well-defined parameter which is constant within $\pm5 \%$ and allows us to infer the location of the HOMO (or LUMO) from the TVS minimum voltage. However, it is usually not well known where the voltage drops for a specific molecule, and this implies an uncertainty about the value of $\chi$: this value will lie somewhere between the upper curve and the lower curve of Fig.~\ref{fig:d-Es-Vm}. Therefore this figure indicates the typical degree of uncertainty one faces in interpreting TVS results when the amount of screening is not known. \\

\subsection{(A)Symmetry}
\label{symmetry}
\begin{figure}
\includegraphics[scale=0.25]{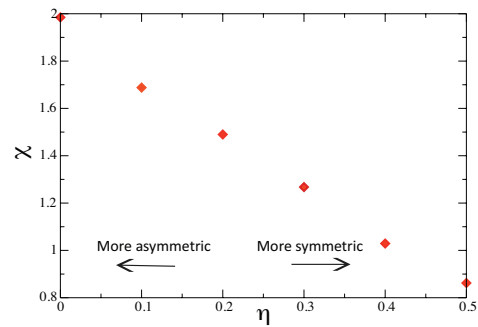}
\caption{\label{fig:sym} $\chi=|E_F-E_s|/V_{\text{min}}$  versus $\eta$ for positive bias voltage and $E_s=0.5$.  \\  }
\end{figure}

\begin{figure}
\includegraphics[scale=0.5]{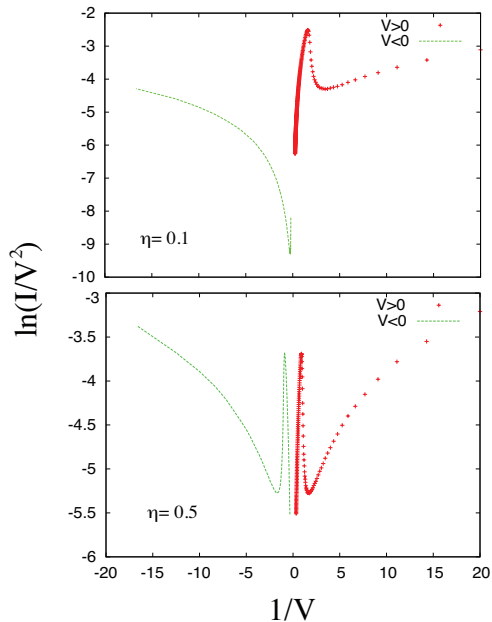}
\caption{\label{fig:eta0.1-0.5} $\ln(I/V^2)$ versus $1/V$ for positive and negative bias voltages in two cases of $\eta=0.1,0.5$. $E_s=0.5$, $t_{L,R}=0.2$ and $U=0$. \\  }
\end{figure}

\begin{figure}
\includegraphics[scale=0.3]{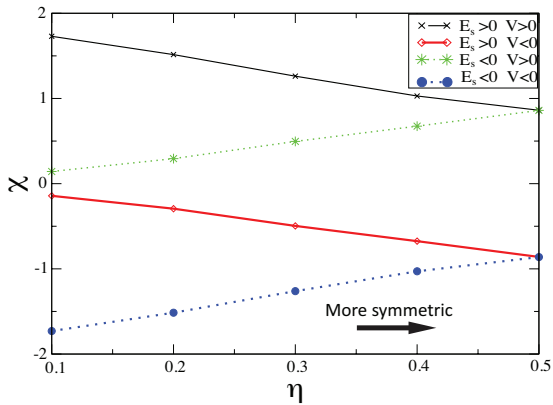}
\caption{\label{fig:posneg} $\chi$ versus $\eta$ for $E_s=\pm 0.5$ and $E_F=0$. The left and right couplings are $t_L=t_R=0.2$. Crosses: $E_s>0$, $V>0$, diamonds: $E_s>0$, $V<0$, stars: $E_s<0$, $V>0$, circles: $E_s<0$,  $V<0$. \\ }
\end{figure}

We now turn to an investigation of the effect of asymmetric capacitive coupling on TVS. This aspect was previously considered in Ref. \cite{Chen}. Here we want to study this using our generic model. The asymmetry is described by a parameter $\eta$, which we define differently from Ref. \cite{Chen}. We change the chemical potential of the contacts by the parameter $\eta$ where $\mu_L=E_F-\eta V$ and $\mu_R=E_F+(1-\eta)V$ \cite{notenote}. In fact, the parameter $\eta$ specifies how the bias voltage is distributed over the left and right contact. In the symmetric case $\eta=1/2$, while in the fully asymmetric case $\eta=0$. Using symmetry $\eta \leftrightarrow 1-\eta$, only $\eta$ between 0 and 0.5 needs to be considered. The ratio of $|E_F-E_s|$ to $V_{\text{min}}$, i.e. $\chi$ versus $\eta$ for positive voltage is shown in Fig.~\ref{fig:sym}. This ratio varies between $0.8$ and $2$, which is in agreement with the result of Ref.~\cite{Chen}. Since $\chi$ depends on symmetry, one can only find the HOMO/ LUMO level energy from $V_{\text{min}}$, if $\eta$ is known. In this curve, the molecular level is reached at $V=|E_F-E_s|/(1-\eta)$. Thus in order to have $V_{\text{min}}$ smaller than the voltage required to reach the level, we should have $\chi > 1-\eta$ which is the case for all of the points in Fig.~\ref{fig:sym}. \\
In Fig.~\ref{fig:eta0.1-0.5}, the FN plot is shown for a symmetric $(\eta=0.5)$ and an asymmetric $(\eta=0.1)$ junction. It can be seen that, for the asymmetric case, the transition voltage differs between positive and negative voltages which is the first important feature that should be taken into account in the case of using TVS.\\
In Fig.~\ref{fig:posneg}, $\chi=|E_F-E_s|/V_{\text{min}}$ is shown as a function of $\eta$ for positive $(\chi_p)$ and for negative voltages $(\chi_n)$ for two different cases : (i) $E_s$ is above the Fermi level $(E_s=0.5 ,E_F=0)$, i.e.\ the resonant level is the LUMO level, and (ii) $E_s$ is below the Fermi level $(E_s=-0.5 ,E_F=0)$, i.e.\ the HOMO level. Here we suppose that only a single level contributes to the transport. The absolute value of $\chi$ is the same for the positive and the negative voltages in the symmetric junction as shown in Fig.~\ref{fig:posneg}. However it is different for the asymmetric case. Moreover, in the case of the asymmetric junction, $|\chi|$ for $E_s>0$, $V<0$ is the same as for $E_s<0$, $V>0$ and also $|\chi|$ for $E_s>0$, $V>0$ is the same as for $E_s<0$, $V<0$.\\
Interestingly, it is possible to estimate the asymmetry of the junction by looking at the ratio $|\chi_{\text{p}}/ \chi_{\text{n}}|=  V_{\text{min,n}}/V_{\text{min,p}}$ where the subscript $p,n$ refers to positive and negative voltages respectively. When $E_s$ is above the Fermi level, the ratio of $|\chi_{\text{p}}/ \chi_{\text{n}}|$ or $V_{\text{min,n}}/V_{\text{min,p}}$  is equal to $(1-\eta)/\eta$. Similarly in the case of $E_s$ below the Fermi level, $|\chi_{\text{p}}/ \chi_{\text{n}}|  $ is equal to $\eta/(1-\eta)$. Hence in principle, the ratio of $V_{\text{min,n}}/V_{\text{min,p}}$ allows us to find the asymmetry degree and, hence remarkably, TVS can be used both for finding the information about the HOMO/LUMO level and about the asymmetry degree of the coupling by looking at positive and negative voltages. In Ref.~\cite{Chen}, an asymmetrically coupled junction was studied for the positive bias only and the authors argued that if the HOMO level can be found by other tools, then TVS can be used to estimate the asymmetry degree or vice versa which is different from our statement. \\
\begin{figure}
\includegraphics[scale=0.25]{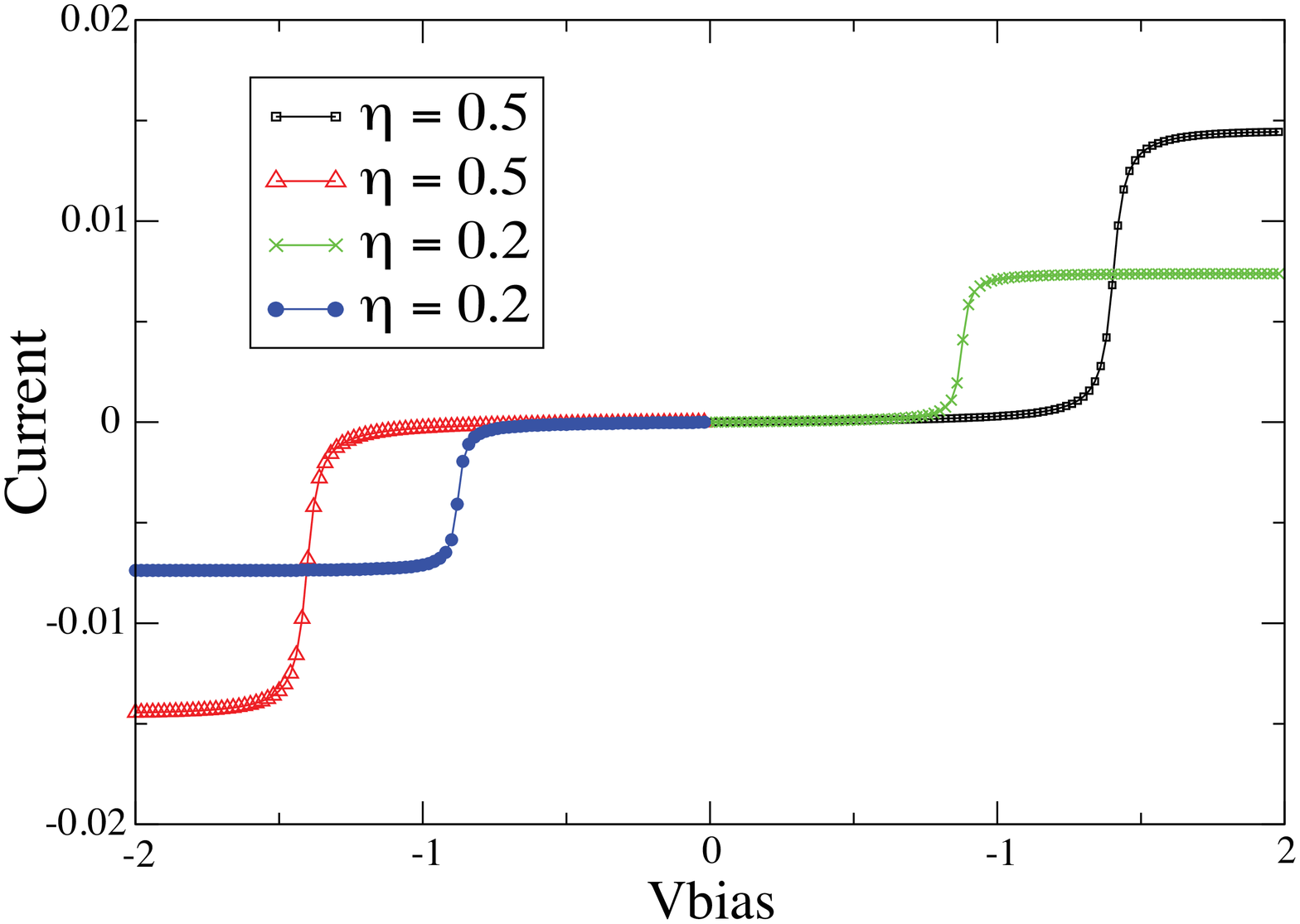}
\caption{\label{fig:IV} Current versus voltage for a two-levels model and $E_{1,2}=\pm 0.7$, $E_F=0$ and $t_{L,R}=0.15$. Squares for positive voltage and $\eta=0.5$. Triangles for negative voltage and $\eta=0.5$. Crosses for positive voltage and $\eta=0.2$. Circles for negative voltage and $\eta=0.2$. \\ }
\end{figure}
Summarizing the results presented in this section, we emphasize that if one depicts the I-V characteristics by a FN plot, both the positive and negative bias voltages should be considered. From that, it is possible to say whether the capacitive coupling is symmetric or asymmetric. By having $V_{\text{min,n}}/V_{\text{min,p}}$ and knowing whether the resonant level is HOMO or LUMO, we showed that it is possible to estimate the degree of asymmetry of the molecular junction. 
However one should note that here also the voltage drop matters. Similar to the discussion in Sec.~\ref{drop}, the lack of information about the voltage drop can lead to an uncertainty in $\chi$ of about $\sim20 \%$ to $30 \% $. 

In some cases, asymmetric coupling may lead to a symmetric I-V. Consider for example a two-level system. The I-V curve for this system is shown in Fig.~\ref{fig:IV}. The two closest energy levels to the Fermi energy $(E_F=0)$, are $E_{1,2}=\pm 0.7$. In this case, even for asymmetric capacitive coupling $(\eta=0.2)$, the I-V curve is symmetric and this is due to the fact that both energy levels are considered in the same distance far from the Fermi energy, with the same coupling. Hence the possibility to use TVS to estimate the asymmetry degree of the molecular junction coupling only exists for the case  where a single level is dominant in the transport. \\
In experiments, usually one level dominates in the transport at low voltages and usually the HOMO and LUMO levels are not in the same distance from the Fermi energy. In the case of a two level model, one could think of HOMO and LUMO energy levels below and above the Fermi energy with different distance from $E_F$. If $|E_F-E_H|> \text{max}(\frac{1-\eta}{\eta},\frac{\eta}{1-\eta}) |E_F-E_L|$ the transport is dominant through the LUMO level. A similar argument leads to dominant HOMO  when $|E_F-E_L|> \text{max}(\frac{1-\eta}{\eta},\frac{\eta}{1-\eta}) |E_F-E_H|$. Both these conditions are the criteria to dominate the transport through one level which provide the possibility to use TVS.

\section{Conclusions}
\label{Conclusion}
In conclusion, we have investigated the length dependence of transition voltage and the influences of (a)symmetric coupling of a molecular junction on TVS. For molecules with strong screening, the HOMO-LUMO gap is usually small compared to alkanethiols and this can explain the relative variation of $V_{\text{min}}$ with $d$ in conjugated molecules. The weak length dependence of $V_{\text{min}}$ in alkanethiols can be elucidated by the strong localization of the orbitals on the end group. Unfortunately in the experiments the lack of information about the potential profile over the molecule limits the usefulness of TVS as a tool for identifying the location of the conducting frontier orbitals. Furthermore we looked at the possibilities and shortcomings of TVS in the case of an asymmetric junction.We showed that in addition to get the information about the HOMO/ LUMO level position by TVS, it is also possible to estimate the degree of asymmetry by looking at the transition voltage value at the positive and negative voltages. It must be noted, however, that also this case is limited by lack of information on the voltage profile inside the molecule. Finally, we showed that in order to estimate the degree of asymmetry by TVS, one should take note of the number of the dominant levels in the transport. Summarizing, TVS may be a useful analysis technique, but it should be used with considerable care. \\

\section{Acknowledgement}
It is a pleasure to thank C. M. Gu{\'e}don for the useful discussions. Financial support was obtained from the EU FP7 program under the grant agreement SINGLE. S.J.v.d.M gratefully acknowledges support from the NWO-VIDI grant. \\

\bibliography{biblio}

\begin{thebibliography}{10}
\bibitem{notenote}%
  \BibitemOpen
  \bibinfo {note} {In the model of Ref [12], the asymmetry degree is contained
  in a parameter $\eta$ which is added to the molecular energy level $(E+\eta
  V)$. In their case, $\eta=0$ corresponds to a symmetric junction and
  $\eta=1/2$ is for the fully asymmetric case.}%
  \bibAnnoteFile{Stop}{notenote}%
\end{thebibliography}
\end{document}